\documentclass[5p,,preprint,12pt]{elsarticle}
\makeatletter\if@twocolumn\PassOptionsToPackage{switch}{lineno}\else\fi\makeatother

\usepackage[utf8]{inputenc} 
\usepackage[T1]{fontenc}    
\usepackage{hyperref}       
\usepackage{url}            
\usepackage{booktabs}       
\usepackage{amsfonts}       
\usepackage{nicefrac}       
\usepackage{microtype}      
\usepackage{lipsum}
\usepackage[numbers]{natbib}
\usepackage{graphicx}
\usepackage{amssymb,amsmath}
\usepackage{mathtools}

\usepackage{tabulary,xcolor}

\DeclarePairedDelimiter{\norm}{\lVert}{\rVert}
	
\makeatletter
\let\save@ps@pprintTitle\ps@pprintTitle
\def\ps@pprintTitle{\save@ps@pprintTitle\gdef\@oddfoot{\footnotesize\itshape \null\hfill\today}}
\def\hlinewd#1{%
  \noalign{\ifnum0=`}\fi\hrule \@height #1%
  \futurelet\reserved@a\@xhline}
\def\tbltoprule{\hlinewd{.8pt}\\[-12pt]}
\def\tblbottomrule{\noalign{\vspace*{6pt}}\hline\noalign{\vspace*{2pt}}}

\AtBeginDocument{\ifNAT@numbers \biboptions{sort&compress}\fi}
\makeatother

\usepackage{ifluatex}
\ifluatex
\usepackage{fontspec}
\defaultfontfeatures{Ligatures=TeX}
\usepackage[]{unicode-math}
\unimathsetup{math-style=TeX}
\else 
\usepackage[utf8]{inputenc}
\fi 
\ifluatex\else\usepackage{stmaryrd}\fi

\usepackage{url,multirow,morefloats,floatflt,cancel,tfrupee}
\makeatletter

\AtBeginDocument{\@ifpackageloaded{textcomp}{}{\usepackage{textcomp}}}
\makeatother
\usepackage{colortbl}
\usepackage{xcolor}
\usepackage{pifont}
\usepackage[nointegrals]{wasysym}
\makeatletter

\def\mcWidth#1{\csname TY@F#1\endcsname+\tabcolsep}

\def\cAlignHack{\rightskip\@flushglue\leftskip\@flushglue\parindent\z@\parfillskip\z@skip}
\def\rAlignHack{\rightskip\z@skip\leftskip\@flushglue \parindent\z@\parfillskip\z@skip}

\@ifundefined{etal}{}{}

\usepackage{ifxetex}
\ifxetex\else\if@twocolumn\@ifpackageloaded{stfloats}{}{\usepackage{dblfloatfix}}\fi\fi

\AtBeginDocument{
\expandafter\ifx\csname eqalign\endcsname\relax
\def\eqalign#1{\null\vcenter{\def\\{\cr}\openup\jot\m@th
  \ialign{\strut$\displaystyle{##}$\hfil&$\displaystyle{{}##}$\hfil
      \crcr#1\crcr}}\,}
\fi
}

\AtBeginDocument{%
  \@ifpackageloaded{endfloat}%
   {\renewcommand\efloat@iwrite[1]{\immediate\expandafter\protected@write\csname efloat@post#1\endcsname{}}}{\newif\ifefloat@tables}%
}%

\def\BreakURLText#1{\@tfor\brk@tempa:=#1\do{\brk@tempa\hskip0pt}}
\let\lt=<
\let\gt=>
\def\processVert{\ifmmode|\else\textbar\fi}

\@ifundefined{subparagraph}{
\def\subparagraph{\@startsection{paragraph}{5}{2\parindent}{0ex plus 0.1ex minus 0.1ex}%
{0ex}{\normalfont\small\itshape}}%
}{}

\newcommand\role[1]{\unskip}
\newcommand\aucollab[1]{\unskip}
  
\@ifundefined{tsGraphicsScaleX}{\gdef\tsGraphicsScaleX{1}}{}
\@ifundefined{tsGraphicsScaleY}{\gdef\tsGraphicsScaleY{.9}}{}
\def\checkGraphicsWidth{\ifdim\Gin@nat@width>\linewidth
	\tsGraphicsScaleX\linewidth\else\Gin@nat@width\fi}

\def\checkGraphicsHeight{\ifdim\Gin@nat@height>.9\textheight
	\tsGraphicsScaleY\textheight\else\Gin@nat@height\fi}

\def\fixFloatSize#1{}
\let\ts@includegraphics\includegraphics

\def\inlinegraphic[#1]#2{{\edef\@tempa{#1}\edef\baseline@shift{\ifx\@tempa\@empty0\else#1\fi}\edef\tempZ{\the\numexpr(\numexpr(\baseline@shift*\f@size/100))}\protect\raisebox{\tempZ pt}{\ts@includegraphics{#2}}}}

\AtBeginDocument{\def\includegraphics{\@ifnextchar[{\ts@includegraphics}{\ts@includegraphics[width=\checkGraphicsWidth,height=\checkGraphicsHeight,keepaspectratio]}}}

\DeclareMathAlphabet{\mathpzc}{OT1}{pzc}{m}{it}

\def\URL#1#2{\@ifundefined{href}{#2}{\href{#1}{#2}}}

\def\UrlOrds{\do\*\do\-\do\~\do\'\do\"\do\-}%
\g@addto@macro{\UrlBreaks}{\UrlOrds}

\edef\fntEncoding{\f@encoding}

\makeatother

\newif\ifmultipleabstract\multipleabstractfalse%
\usepackage{float}

\begin{document}

\begin{frontmatter}
	
\title{Correlation Distance Skip Connection Denoising Autoencoder (CDSK-DAE) for Speech Feature Enhancement
}
    
\author[ace115e830958]{Alzahra Badi}
\author[aa9e62b49529f]{Sangwook Park}
\author[ab913344b17cc]{David K. Han}
\author[ace115e830958]{Hanseok Ko\corref{c-5bf4d895f9c2}}
\ead{hsko@korea.ac.kr}\cortext[c-5bf4d895f9c2]{Corresponding author.}
    
\address[ace115e830958]{Electrical and Computer Science Engineering\unskip, 
    Korea University}
  	
\address[aa9e62b49529f]{Electrical and Computer Engineering\unskip, 
    Johns Hopkins University}
  	
\address[ab913344b17cc]{Information Sciences Division\unskip, 
    U.S. Army Research Laboratory}

\begin{abstract}
Performance of learning based Automatic Speech Recognition (ASR) is susceptible to noise, especially when it is introduced in the testing data while not presented in the training data. This work focuses on a feature enhancement for noise robust end-to-end ASR system by introducing a novel variant of denoising autoencoder (DAE). The proposed method uses skip connections in both encoder and decoder sides by passing speech information of the target frame from input to the model. It also uses a new objective function in training model that uses a correlation distance measure in penalty terms by measuring dependency of the latent target features and the model (latent features and enhanced features obtained from the DAE). Performance of the proposed method was compared against a conventional model and a state of the art model under both seen and unseen noisy environments of 7 different types of background noise with different SNR levels (0, 5, 10 and 20 dB). The proposed method also is tested using linear and non-linear penalty terms as well, where, they both show an improvement on the overall average WER under noisy conditions both seen and unseen in comparison to the state-of-the-art model. 

\end{abstract}
      
\end{frontmatter}
    
\section{Introduction}
Automatic Speech Recognition (ASR) field has been a research area focused on generating text from human speech. Traditionally, ASR system requires integrating different stages of models: namely acoustic, language and pronunciation models. An acoustic model first takes raw audio input and develop statistical models via extracting acoustic features such as Mel Frequency Cepstral Coefficients. These features are then applied to an observation model of the Hidden Markov Model (HMM) for temporal variability to estimate text sequence of the audio stream input. HMM integrates in its process the Language Model (LM) typically constructed of word sequence probabilities given a phone sequence \cite{Goodman2001}.

Over the last decade, there has been a dramatic shift in the ASR community due to the explosion of deep learning. Although the idea of deep layered network has been around over many decades, the sudden embrace of the method by the ASR community can be attributed to availability of both computational power and large public datasets that would allow training of deep layered networks possible.  Effectiveness of deep networks has been such that traditionally handcrafted feature based methods, such as GMM, was mostly replaced by the new paradigm. The approach yielded a significant improvement in performance of ASR systems \cite{Dahl2012a,Mohamed2012,Mohamed2009}.  Application of deep network to other areas such as lexicon and language followed for improved performance \cite{Rao2015,Mikolov2011,Xu2018,Sundermeyer2012}. In these earlier implementations of deep learning into ASR, the structures still require integration of these components mentioned.

Recently, end-to-end models for ASR have been an active research focus. These end-to-end models reduce the complexity of the system by making a direct mapping of input speech frames to output texts without the need of intermediate model such as GMM, HMM, or LM \cite{Hannun2014,Graves2014,Zhang2017b,Miao2015}. These end-to-end models, such as sequence-to-sequence \cite{Zhang2017b,Chan2016a,Bahdanau2015} or Connectionist Temporal Classification (CTC) based ASR \cite{Graves2014,Miao2015,Graves2006}, have achieved good performance in comparison to the conventional ASR models. These models reduce the complexity of deep learning based ASR systems significantly in terms of their structures as well as the training of the networks, and also delivers good performance.

As in most of the ASR approaches, including the ones mentioned here so far, however, the intelligibility of an ASR system is heavily influenced by presence of noise. While building an effective ASR system is a challenge, building an ASR system that is robust to noise remains more difficult. For traditional hand-crafted feature based systems, MFCC and log Mel-filter bank features are used as input to the model because they are based on the human auditory system. However, these features have no strategies for preventing the performance degradation in noisy conditions. For handling noise using deep learning architectures, there have been a variety of methods proposed including denoising autoencoder, which uses a non-linear mapping of noisy speech to target clean speech features. Deep Denoising Autoencoder (DDAE) \cite{Lu2013a,Feng2014a}, Denoising Variational Autoencoder (VAE) \cite{Hsu2018}, and multi-tasking autoencoder (MTAE) \cite{Zhang2018b} are some of these autoencoder based approaches applied to alleviate the noise issue.

We believe that ASR performance is highly dependent on how well noise problem is handled upfront to prevent corruption of signals for downstream chain. Thus, we propose a novel noise robust ASR approach with the following components. Within the front-end of the ASR framework, this work focuses on the feature enhancement method by appending the pre-processing stage by implementing a denoising autoencoder with skip connections where the original input is fed into different stages in the network. In addition, a correlation distance measure (CDM) that has been proposed in \cite{Szekely2007} and had been used in clustering \cite{JyotiBora2014} with better demonstrated performance over other similarity distance measures, is implemented.  The CDM is used in the objective function to increase the dependency between the latent and enhanced with the target features. The performance of our approach is measured by WER that is obtained by training a Recurrent Neural Network (RNN)-CTC phoneme-based ASR model using the EESEN toolkit .  We used DDAE \citep{Lu2013a}, and MTAE \cite{Zhang2018b} as baseline methods for comparison purpose.

\begin{figure*}[hbt!]
	\centering
	\includegraphics[width=18cm, height=7cm]{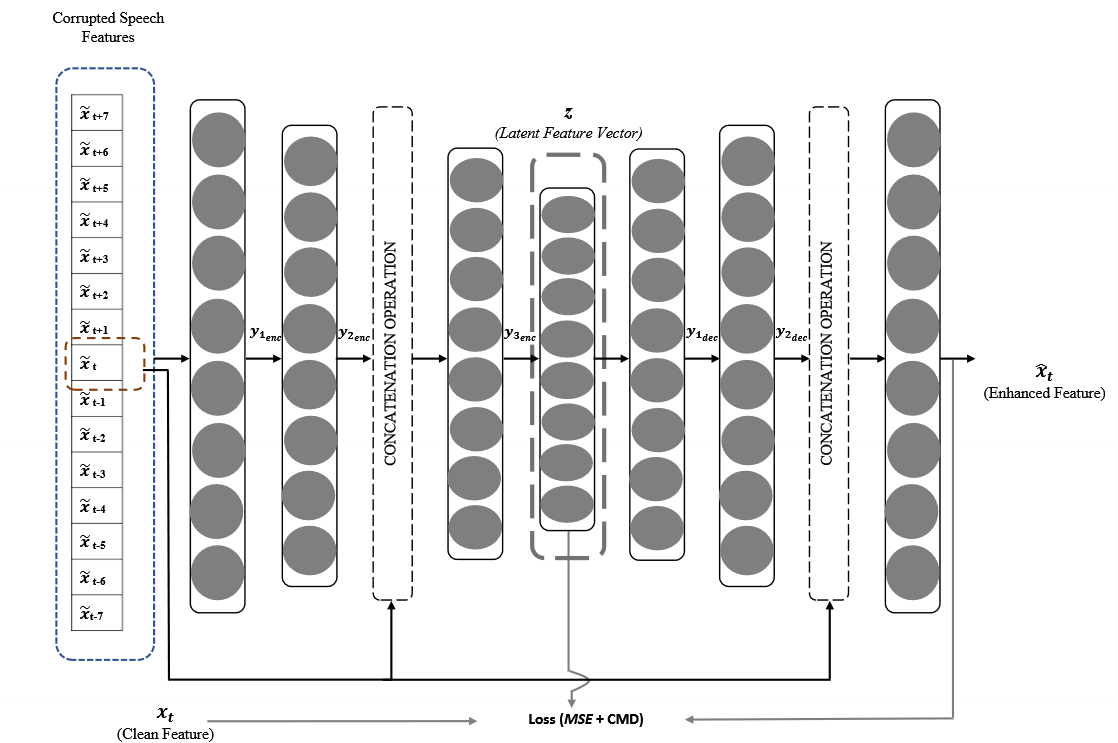}
	\caption{Skip-connection denoising autoencoder block diagram structure }
	\label{fig:1}       
\end{figure*}

\section{Related Work}

\subsection{Denoising Autoencoder (DAE)} A denoising autoencoder, a variant of basic autoencoder, consists of two parts: the encoder and the decoder. The DAE is trained to reconstruct clean target speech from corrupted input. The encoder maps the input $\tilde{x} $ to a hidden representation $z=f(W\tilde{x}+b) $, and the decoder reconstructs an enhanced version $\hat{x}=f(Wz+b) $, where f denotes affine transformation \cite{Vincent2008}. This is done by minimizing the loss, computed from the mean squared error ($\textit{MSE} $) between the clean target ($\textit{x} $) and the output enhanced version of the model ($\hat{x}$),  as in Equation~(\ref{eq:1}). 

\begin{equation}
\label{eq:1}
Loss(x,\hat{x})=||x-\hat{x}||^2
\end{equation} 

where $\widetilde x $ corresponds to the corrupted speech features, $\hat{x}$ is the enhanced speech features, and x is the ground truth speech features \cite{Lu2013a,Feng2014a,Lu2012}. In a typical implementation, the speech features input to the DAE is not in the same dimension of the clean target speech features. This is because adjacent frames are also added as input by concatenating them with the current frame with knowledge that there is often leakage of speech information among neighbouring frames. Assuming noise being IID, however, contribution of noise from neighbouring frames should be minimal. More extended version of DAE has been proposed such as deep denoising autoencoder \cite{Lu2013a} with 5 hidden layers and using RBM training, and Multitasking Autoencoder where denoising task and noise features estimation task are performed separately in the DAE with mutually shared units among the two tasks \cite{Zhang2018b}.

\subsection{Skip connection networks} The skip connection, also known as shortcut connection, has been employed in image recognition and achieved successful results as in highway networks \cite{Srivastava2015,He2016b}.  It is also deployed in various image applications ranging from classification to image restoration by using autoencoder \cite{Mao2016}. In addition, the skip connection has been also used in acoustic field with applications in denoising autoencoder for music \cite{Liu2018}, speech enhancement \cite{Tu2017}, and others.

Additionally, different types of skip connection in image processing were investigated \cite{He2016} where, the identity shortcut had given the best performance. This identity shortcut has no extra parameters to be learned so it reduces the complexity and ensures information flow. Based on that, our proposed model is using an identity skip connection in the proposed structure.

\section{Proposed Model}
The proposed method is composed of two main parts: a novel skip connection architecture and a new objective function shown to be effective in noise removal. Novelty of our approach is an implementation of skip connections to the denoising autoencoder structure with an addition of a correlation distance measure in the objective function to enhance contribution of information from the skip connection for improved speech feature reconstruction in ASR. Figure~\ref{fig:1} illustrates the proposed skip-connection structure.

\subsection{Skip Connection Denoising Autoencoder (SK-DDAE) } \paragraph{Encoder} \sloppy The encoder extracts features of an input to a latent space by finding the best representation of the input with minimal influence by noise.  Input $\tilde{X}$ is fed to the proposed structure consisting of 11 concatenated frames of log Mel-filter bank ($\tilde{X}  = [\tilde{x}_{t-5},\tilde{x}_{t-4}, \dotsb ,\tilde{x}_{t-1},\tilde{x}_t,\tilde{x}_{t+1}, \dotsb, \tilde{x}_{t+4}, \tilde{x}_{t+5}] $) normalized in the range of [0,1] per utterance. We implement an identity skip connection in the encoder to allow the target frame, ($\tilde{x_t}$), to be fed directly into the middle layer of the encoder. The information of the skip connection ($\tilde{x_t}$) is concatenating with the output from the previous layer ($y_{i_enc-1}$), to be fed to the next layer of the encoder (i\textsuperscript{th} layer), and the output can be obtained as in Equation~(\ref{eq:2}).
\begin{equation}
\label{eq:2}
y_{i_{enc}}=\sigma(W.[y_{i_{enc}-1)},\tilde{x}_t ]+b)
\end{equation}

where $W$, $b$ and $\sigma$ are the weight, bias and the sigmoid function, respectively. The skip connection in our proposed model is somewhat different than the residual learning that was introduced in \citep{He2016b,Veit2016}. While their skip connection mechanism adds the input frame information, we instead concatenate the input frame $\tilde{x_t}$ to the mid-layer. We believe this approach improves the network to find correlated information in the required frame and carry it to influence the output for better representation in the output. The encoder in the proposed model consists of three layers of MLP with 512, 256 and 128 units. 

\paragraph{Decoder} Purpose of the decoder is to reconstruct the target frame features from the latent features obtained from the encoder. The same skip connection (shortcut) from the input is also used as part of the decoder with the input frame concatenated to a mid-layer in the decoder based on the same premise that additional correlated information from the input frame can be gleaned in this stage. The decoder layers are organized in a symmetric manner to the encoder layers with 128, 256 and 512 used units, and finally only 40 enhanced features of the desired frame ($\hat{x}$) is obtained. The obtained features, eventually used for our ASR model, resulted better performance when noise is present in the input.

\subsection{Objective function}
In this paper, we are proposing a new objective function by adding a penalty term to the reconstruction error. The penalty is based on the Correlation Distance matrix approach that has been proposed in \cite{Szekely2007}. The correlation distance matrix had been used for clustering purpose as it is a similarity measure that considers as a true measure of dependency and test the joint independence between two random vectors. The dependency measure would add a new constraint in the network training to force the network to extract more salient features relevant to ASR.

\paragraph{Correlation Distance}  Assuming two random samples $(X,Y)={(X_k,Y_k ): k=1,\dotsb
, n} $ from a joint distribution of X in ${\mathbb{R}}^{n}$ and Y in ${\mathbb{R}}^{q}$, we compute the correlation  distance measure($\mathcal{\Re}(X,Y)$) as in Equation~(\ref{eq:3}).

\begin{equation}
\label{eq:3}
\begin{aligned}
\mathcal{\Re}^2(X,Y)= 
\begin{cases}
\frac{\mathcal{V}^2(X,Y)}{\sqrt[2]{\mathcal{V}^2(X)\mathcal{V}^2(Y)}}, 	&\text{$\mathcal{V}^2(X)\mathcal{V}^2(Y)>0$,}\\

0,		&\text{$\mathcal{V}^2(X)\mathcal{V}^2(Y)=0$}
\end{cases}
\end{aligned}
\end{equation}

where,  $V^{2}(X,Y) $is the covariance distance that is obtained by computing the double centered distance matrices A and B by defining:

\begin{equation}
\label{eq:4}
\begin{aligned}
a_{kl}=|X_k- X_l|_p ,	\bar{a}_{k.}=1/n\sum_{l=1}^{n} a_{kl} ,\\
	\bar{a}_{.l}=1/n\sum_{k=1}^{n}a_{kl} ,	\bar{a}_{..}=1/n^2 \sum_{k,l=1}^n a_{kl}
	\end{aligned}
\end{equation}
\begin{equation}
\label{eq:5}
A_{kl}=a_{kl}-\bar{a}_{k.}-\bar{a}_{.l}+\bar{a}_{..}
\end{equation}

Similarly, 
\begin{equation}
\begin{aligned}
\label{eq:6}
b_{kl}=|Y_k- Y_l |_p ,	\bar{b}_{k.}=1/n \sum_{l=1}^nb_{kl} ,\\
	\bar{b}_{.l}=1/n\sum_{k=1}^nb_{kl} ,	\bar{b}_{..}=1/n^2\sum_{(k,l=1)}^{n}b_{kl}
\end{aligned}
\end{equation}

\begin{equation}
\label{eq:7}
B_{kl}=b_{kl}-\bar{b}_{k.}-\bar{b}_{.l}+\bar{b}_{..}
\end{equation}

Hence,

\begin{equation}
\label{eq:8}
\begin{aligned}
V^2 (X,Y)=1/n^2\sum_{(k,l=1)}^{n}{A_{kl} B_{kl}}
\end{aligned}
\end{equation}

The  $\mathcal{\Re}(X,Y) $ guarantees nonnegativity and it is in range of [0,1] in contrast to the Pearson correlation coefficient that has a range of [-1,1]. Additionally, when  $\mathcal{\Re}(X,Y) = 0 $, it indicates independence between these two random vectors. However, when $\mathcal{\Re}(X,Y) = 1 $, it indicates linear dependency \cite{Vincent2008}.

Therefore, the objective function that is used for training the SK-DDAE can be expressed as:

\begin{equation}
\begin{aligned}
\label{eq:9}
loss=\norm{x_t-\hat{x}}_2^2+\beta*((1-\mathcal{\Re}(z,x_t))+\\
(1-\mathcal{\Re}(\hat{x},x_t)))
\end{aligned}
\end{equation}

  where, $z$ and $\beta$ are the latent features that are obtained from the encoder part and penalty term parameter, respectively. The objective function will maximize the dependency between the latent features and the target, and would drive the network toward producing $z$ that has good representation of the target by minimizing noise information. Additionally, it also measures the dependency between the output and target, and try to increase the dependency in the training process. Furthermore, a second variation of the proposed model was investigated by using a nonlinear term in the objective function (Energy term) that we assume will enhance the system performance as:

\begin{multline}
\label{eq:10}
loss=\norm{x_t-\hat{x}}_2^2+\beta*(1-\mathcal{\Re}(z,x_t )+(1-\mathcal{\Re}(\hat{x},x_t ))\\
+\sigma*((1-\mathcal{\Re}(z,x_t))^2+(1-\mathcal{\Re}(\hat{x},x_t))^2
\end{multline}

In this paper, we will refer the model that uses only the mean squared error loss as SK-DAE, the linear penalty correlation distance term as CDSK-DAE, and the nonlinear penalty correlation distance term as CDESK-DAE.

In addition, we use $\beta$, $\sigma $ = 0.01\$ for simplicity, which we leave the process of finding the optimal penalty term parameters ($\beta$ and $\sigma$) for future work. Also, we found that the structure performs better with identity skip connection than using CNN skip connection which is consistent with the finding in \cite{He2016}.

\begin{table*}[!htbp]
\caption{{Results comparison by using word error rate (WER\%) per method in clean condition} }
\label{1}
\def\arraystretch{1}
\ignorespaces 
\centering 
\begin{tabulary}{\linewidth}{LLLLLL}
\tbltoprule 
Log Mel-Filter Bank &  MTAE (WER\%) &  DDAE (WER\%) &  SK-DDAE (WER\%)  &  CDESK-DAE (WER\%)  &  CDSK-DAE (WER\%) \\
7.92 &  9.85 &  10.01 &  9.54  &  8.49  &   8.65 \\
\tblbottomrule 
\end{tabulary}\par 
\end{table*}

\begin{table*}[!htbp]
\caption{{Results comparison by using word error rate percentage (WER\%) between the baselines and proposed model against different type of seen noise and levels of SNR} }
\label{2}
\def\arraystretch{1}
\ignorespaces 
\centering 
\begin{tabulary}{\linewidth}{LLLLLLLL}
\tbltoprule 
SNR (dB)  &  Mel Filter Bank (WER\%)  &  MTAE (WER\%)  &  DDAE (WER\%)  &  SK-DDAE (WER\%)  &  CDESK-DAE (WER\%) &  CDSK-DAE (WER\%) &  \\
\multicolumn{8}{p{\dimexpr(\mcWidth{1}+\mcWidth{2}+\mcWidth{3}+\mcWidth{4}+\mcWidth{5}+\mcWidth{6}+\mcWidth{7}+\mcWidth{8})}}{Shopping center }\\
0 &  55.18 &  40.14 &  44.59 &  39.09 &  37.46 &  38.53 &  \\
5 &  35.58 &  20.96 &  28.5 &  25.5 &  25.59 &  27.7 &  \\
  10 &  20.27 &  23.27 &  19.74 &  18.16 &  18.68 &  19.32 &  \\
  20 &  10.81 &  15.15 &  13.59 &  12.62 &  12.51 &  12.92 &  \\
\multicolumn{8}{p{\dimexpr(\mcWidth{1}+\mcWidth{2}+\mcWidth{3}+\mcWidth{4}+\mcWidth{5}+\mcWidth{6}+\mcWidth{7}+\mcWidth{8})}}{Pub}\\
0 &  90.31 &  65.53 &  78.77 &  70.39 &  71.54 &  69.36 &  \\
5 &  60.13 &  48.56 &  51.02 &  46.73 &  46.59 &  45.74 &  \\
10 &  31.84 &  32.84 &  30.52 &  29.24 &  28.51 &  28.65 &  \\
20 &  11.84 &  17.15 &  15.13 &  14.71 &  14.67 &  13.96 &  \\
\multicolumn{8}{p{\dimexpr(\mcWidth{1}+\mcWidth{2}+\mcWidth{3}+\mcWidth{4}+\mcWidth{5}+\mcWidth{6}+\mcWidth{7}+\mcWidth{8})}}{Caf{\'e}}\\
0 &  74.87 &  56.09 &  62.98 &  55.18 &  54.03 &  55.38 &  \\
5 &  51.73 &  41.66 &  43.95 &  39.04 &  37.34 &  40.17 &  \\
10 &  30 &  32.27 &  29.72 &  27.27 &  27.04 &  29.22 &  \\
20 &  11.78 &  18.08 &  15.88 &  14.55 &  15.35 &  15.01 &  \\
\multicolumn{8}{p{\dimexpr(\mcWidth{1}+\mcWidth{2}+\mcWidth{3}+\mcWidth{4}+\mcWidth{5}+\mcWidth{6}+\mcWidth{7}+\mcWidth{8})}}{Schoolyard}\\
0 &  82.79 &  61.79 &  66.77 &  62.61 &  59.4 &  60.22 &  \\
5 &  56.23 &  43.15 &  42.71 &  39.06 &  36.75 &  38.88 &  \\
10 &  33.19 &  31.35 &  25.66 &  25.5 &  24.76 &  27.08 &  \\
20 &  11.93 &  17.17 &  13.72 &  13.91 &  13.91 &  13.2 &  \\
\tblbottomrule 
\end{tabulary}\par 
\end{table*}

\section{Experiment}
The experiment is conducted in two stages; first one is about training the skip- connection denoising autoencoder to obtain the enhanced features. In the second stage, the enhanced features are used in the CTC ASR model based on the EESEN to evaluate the performance of the proposed method.

\subsection{Dataset}The skip-connection autoencoder is trained using the TIMIT dataset, which uses 3696 utterances of clean speech and 14784 augmented utterances with background noise.  The TIMIT dataset (3696 utterance) contains 3.14 hours of data and it is an Acoustic- Phonetic Continuous Speech Corpus, which has 61 phonemes in total. The TIMIT utterances were augmented with 4 types of noise (caf{\'e}, pub, schoolyard, shopping center) with four different SNR levels (0dB, 5dB, 10dB, 20dB). On the other hand, the EESEN ASR model was trained on the WSJ's si-84 data subset which was divided into 95\% for training and 5\% for validation. For evaluation, the WSJ test set (Nov 92) data is augmented with seven type of seen and unseen noise for the SK-DDAE (caf{\'e}, pub, schoolyard, shopping center, living room, nature creek, outside traffic road) obtained from the ETSI background noise database \cite{EuropeanTelecommunicationsStandardsInstitute2008} using ADDNOISE MATLAB library \cite{ITU-T2011a}. In the evaluation, background noises had been added with four different SNR level.s (0dB, 5dB, 10dB, 20dB). Both the WSJ and TIMIT data are obtained from Linguistic Data Consortium (LDC).

\subsection{Experimental setting}The log Mel-filter bank features were extracted with 512 FFT bins with 40 filterbanks to obtain 40 dimensional features. These features are normalized using CMVN and then the delta-delta is added to create a total of 120 features to train the acoustic model.  Furthermore, the denoising autoencoder baselines (DDAE and MTAE) and the proposed methods, are trained using normalized log Mel-filterbank features with batch size of 500. Both the DDAE and MTAE are trained with 50 epochs, whereas, the SK-DDAE is trained with 16 epochs. The proposed model is initialized using the Xavier initialization \cite{Jia2014} and uses AdamOptimizer with learning rate of 0.001. For Evaluation, normalized 40 dimensional features per frame was obtained for the WSJ dataset. A delta-delta features are added to the enhanced features to obtain 120 dimensions that are used for the acoustic model. Furthermore, the acoustic model is a RNN-CTC phoneme based model that have been developed using the EESEN toolkit and uses the same configuration as described in \cite{Miao2015}. The RNN is 4 bidirectional LSTM layers with 320 memory cells for both forward and backward sub-layers.

\begin{table*}[!htbp]
\caption{{Results comparison by using word error rate percentage (WER\%) between the deep learning baselines and proposed model against different type of unseen noise and levels of SNR } }
\label{3}
\def\arraystretch{1}
\ignorespaces 
\centering 
\begin{tabulary}{\linewidth}{LLLLLL}
\tbltoprule 
SNR (dB) &
  MTAE (WER\%) &
  DDAE (WER\%) &
  SK-DDAE (WER\%) &
  CDESK-DAE (WER\%) &
  CDSK-DAE (WER\%)\\
\multicolumn{6}{p{\dimexpr(\mcWidth{1}+\mcWidth{2}+\mcWidth{3}+\mcWidth{4}+\mcWidth{5}+\mcWidth{6})}}{Livingroom }\\
0 &  60.3 &  53.41 &  50.77 &  50.54 &  49.78\\
5 &  42.94 &  36.43 &  35.87 &  34.95 &  35.25\\
10 &  30.11 &  24.03 &  24.26 &  23.87 &  25.31\\
20 &  15.97 &  14.21 &  13.37 &  13.29 &  13.63\\
\multicolumn{6}{p{\dimexpr(\mcWidth{1}+\mcWidth{2}+\mcWidth{3}+\mcWidth{4}+\mcWidth{5}+\mcWidth{6})}}{Natural Creek }\\
0 &  95.16 &  92.79 &  87.17 &  88.06 &  86.85\\
5 &  70.88 &  70.81 &  65.98 &  62.93 &  63.76\\
10 &  45.83 &  43.03 &  40.44 &  38.6 &  40.23\\
20 &  19.74 &  16.06 &  15.77 &  16.27 &  15.59\\
\multicolumn{6}{p{\dimexpr(\mcWidth{1}+\mcWidth{2}+\mcWidth{3}+\mcWidth{4}+\mcWidth{5}+\mcWidth{6})}}{Outside Traffic Road }\\
0 &  67.91 &  66.12 &  61.01 &  60.62 &  60.77\\
5 &  47.26 &  44.21 &  40.44 &  41.02 &  41.36\\
10 &  32.02 &  26.72 &  24.21 &  24.3 &  25.22\\
20 &  15.9 &  13.98 &  13.7 &  13.88 &  13.66\\
\tblbottomrule 
\end{tabulary}\par 
\end{table*}

\begin{table*}[!htbp]
\caption{{Correlation distance measure among the noise and speech utterance in TIMIT training data} }
\label{4}
\def\arraystretch{1}
\ignorespaces 
\centering 
\begin{tabulary}{\linewidth}{p{4.5em}p{4.5em}p{4.5em}p{4.5em}p{4.5em}p{4.5em}p{4.5em}p{4.5em}p{4.5em}}
\tbltoprule 
    & Shopping Center &  Caf{\'e} &  Pub &  Schoolyard  &
  Livingroom  &  Natural Creek &  Outside Traffic Road\\
Speech &  0.1473	& 0.1441	& 0.1781	& 0.1897	& 0.1419	& 0.1435  &	0.1699 \\
\tblbottomrule 
\end{tabulary}\par 
\end{table*}

\begin{table}[!htbp]
\caption{{Correlation distance measure among the training and testing background noise} }
\label{5}
\def\arraystretch{1}
\ignorespaces 
\centering 
\begin{tabulary}{\linewidth}{p{4.3em}p{4.3em}p{4.3em}p{4.3em}p{4.3em}}
\tbltoprule 
Noise Type	& Livingroom	& Natural Creak	& Outside Traffic Road\\
Shopping Center & 0.2161 & 0.2249 & 0.2542\\
Caf{\'e} & 0.2042 & 0.1850 & 0.2664\\
Pub & 0.2271 & 0.2439 & 0.2576\\
Schoolyard & 0.2597 & 0.2515 & 0.3041\\
\tblbottomrule 
\end{tabulary}\par 
\end{table}
    
\section{Results and Discussion}
The proposed model was tested in three different sets of values for $\beta$ and $\sigma$ ($\beta$=0, $\sigma$=0 for SK-DDAE, $\beta$=0.01 and $\sigma$=0 for  CDSK-DAE, and both $\beta$, $\sigma$=0.01 for CDESK-DAE). The result on Table~\ref*{1} illustrates the word error rate on the WSJ database in a clean environment. Both CDESK-DAE and CDSK-DAE performed better than the other deep learning methods (DDAE and MTAE) in identical conditions. 

In both Tables~\ref*{2} and~\ref*{3} , the results illustrate the performance sensitivity of the baseline and the proposed model to various levels of SNR, as well the types of noise, both seen and unseen, in Word Error Rate (WER). Based on these results, the proposed models (CDSK-DAE and CDESK-DAE) outperformed the baseline models (MTAE and DDAE) under all noise types and under all SNR for seen and unseen noises except for very few cases. The performance improvement rates of CDESK-DAE ranging from 7.25\% to 10.14\% for seen noises and 13.91\% and 6.67\% for unseen noise were obtained when the performance of MTAE and DDAE were used as baselines. Similarly, CDSK-DDAE demonstrated  improvement rates ranging 5.27\% to 8.21\% for seen noise and 13.35\% to 6.06\% for unseen noise. Even SK-DDAE which is only trained with mean squared error showed improvement rates in comparison to MTAE and DDAE for seen noise ranging 5.59\% to 8.52\% and unseen noise ranging 13.06\% to 5.74\%, respectively. More significantly, the proposed models have demonstrated generalization capability by showing significant performance improvements of the WER\% in the cases of the unseen noises.  

\begin{figure}
	\centering
	\includegraphics{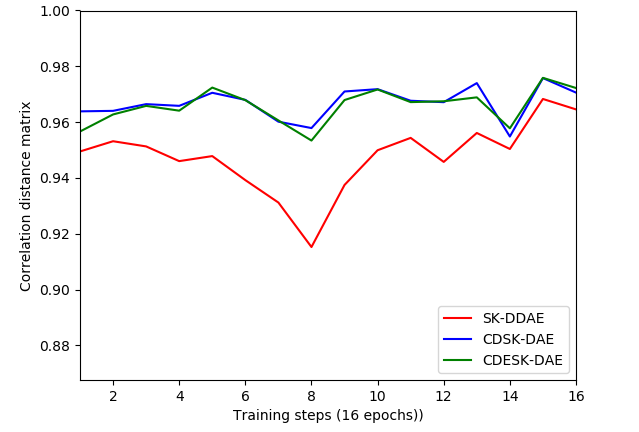}
	\vspace*{-5mm}
	\caption{Distance correlation matrix during training between the latent features ($z$) and target clean features ($x$)}
	\label{fig:2}       
\end{figure}

\begin{figure}
	\centering
	\includegraphics{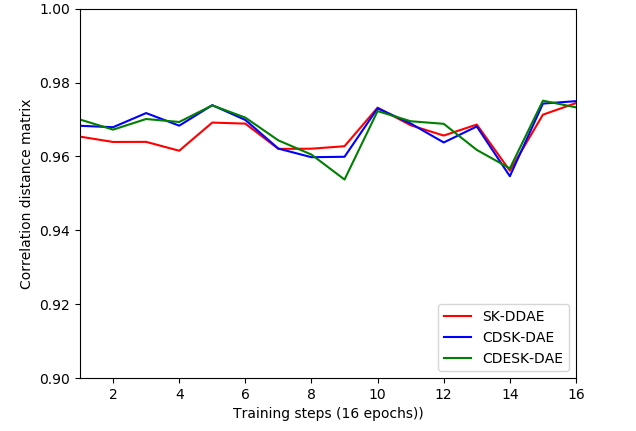}
	\vspace*{-5mm}
	\caption{Distance correlation matrix during training between the enhanced features ($\hat{x}$) and the target clean features ($x$)}
	\label{fig:3}       
\end{figure}

Figure \ref{fig:2} demonstrates the correlation distance matrix of the models (SK-DAE, CDMSK-DAE and CDMESK-DAE) during the training between the latent features with the target features. The addition correlation distance measure in the objective function increases the dependency or relation between the extracted latent features and the target features as shown in the CDMSK-DAE and CDMESK-DAE, in contrast to the SK-DDAE. This increase of dependency results in obtaining a better representation of our target by extracted relevant features in the latent space that helps in the reconstruction process. Similarly, Figure \ref{fig:3} shows that the correlation distance matrix of these three models during the training between the obtained enhanced features ($\hat{x}$) and the target features ($x$). Where the proposed objective function in both CDSK-DAE and CDESK-DAE gives slightly stable high dependency that is maintained through the training than the SK-DDAE in the first epochs.  

Furthermore, Table \ref{4} illustrates the dependency among both the speech utterance (TIMIT train dataset) and the noise type that is used in this work using distance correlation measure. Pub, schoolyard and outside traffic road background noise have the highest correlation distance measure and that is reflected also from the high obtained WER\% from the ASR model. Although, natural creek background noise was not much correlated with the speech, it still gives the highest WER\%. That is because it was less correlated with the training background noise as illustrated in Table \ref{5}.

In summary, the results show that the proposed model with the non-linear correlation distance measure penalty (CDESK-DAE) has the best overall performance under seen and unseen noisy environment. Additionally, the obtained results demonstrate effectiveness of the objective function that uses the correlation distance measure. As assumed, the correlation distance measure seems to increase such dependency between the model and the target during the training, which resulted in better enhanced features that can be used for the ASR system. Although the newly introduced objective function clearly demonstrates its effectiveness, we believe that further research could prompt additional improvements for the denoising task in the ASR applications and beyond.

\section{Conclusions}
In this paper, we proposed a novel architecture of denoising autoencoder by developing a new objective function that uses a correlation distance measure to enhance the dependency between the model and the target features. The model was able to enhance the corrupted features with noise and showed an improvement of performance in both seen and unseen noisy environments. Although the models were effective, we believe further performance improvement can be achieved by optimizing the parameters $\beta$ and $\sigma$.  Additionally, we will investigate the objective function's effectiveness in other acoustic applications.

\section*{Acknowledgments}The work has been supported by the Korea Environmental Industry and Technology Institute's environmental policy-based public technology development project (2017000210001) that is funded by the Ministry of Environment and the contribution of David was supported by the US Army Research Laboratory.

\bibliographystyle{vancouver}

\bibliography{CDSK-DAE}

\end{document}